\documentclass[aps,prd,reprint,amsmath,amssymb,showpacs,superscriptaddress]{revtex4-1}
\usepackage{graphicx}
\usepackage{subfigure}
\usepackage{dcolumn}
\usepackage{bm}
\usepackage{slashed}
\usepackage{ulem}
\usepackage{MnSymbol}
\usepackage{epstopdf}
\usepackage{mathrsfs}
\usepackage{color,xcolor}

\usepackage{hyperref}

\begin{document}
\title{ Flavor dependence of the thermal dissociations of vector and axial-vector mesons}

\author{Ling-Feng Chen}
\affiliation{Department of Physics and State Key Laboratory of Nuclear Physics and Technology, Peking University, Beijing 100871, China.}

\author{Si-Xue Qin}
\email{sqin@cqu.edu.cn}
\affiliation{Department of Physics, Chongqing University, Chongqing 401331, P. R. China.}

\author{Yu-Xin Liu}
\email{yxliu@pku.edu.cn}
\affiliation{Department of Physics and State Key Laboratory of Nuclear Physics and Technology, Peking University, Beijing 100871, China.}
\affiliation{Collaborative Innovation Center of Quantum Matter, Beijing 100871, China.}
\affiliation{Center for High Energy Physics, Peking University, Beijing 100871, China.}

\date{\today}

\begin{abstract}
The in-medium behavior of ground-state $q\bar{q}$ mesons, where $q \in \{u,d,s,c\}$, in vector and axial-vector channels is studied based on the spectral analysis for mesonic correlators at finite temperature and zero chemical potential. We first compute the correlators by solving the quark gap equations and the inhomogeneous Bethe-Salpeter equations in the rainbow-ladder approximation. Using a phenomenological ansatz, the spectral functions are extracted by fitting the correlators. By analyzing the evolution of the spectral functions with the temperature, we obtain the dissociation temperatures of mesons and discuss their relations to the critical temperature of the chiral symmetry restoration. The results show a pattern of the flavor dependence of the thermal dissociation of the mesons.
\end{abstract}

\maketitle

\section{Introduction}
\label{intro}
It is expected that the strongly interacting matter undergoes a crossover transition at a sufficiently high temperature,
above which the dynamically broken chiral symmetry is restored and quarks and gluons are deconfined.
At high temperature, the fundamental degrees of freedom, i.e., quarks and gluons, may form a novel quark-gluon plasma (QGP) state.
The evidence of QGP is accessible by analyzing various indirect hadronic and leptonic signals at the Relativistic Heavy Ion Collider (RHIC)
and the Large Hadron Collider (LHC)~\cite{phenix:2008jexp,sci:2012}.

Heavy quarkonia, i.e., bound states of heavy flavor quarks and antiquarks, are believed to be an ideal probe for the QGP formation~\cite{Aarts:2017}.
For instance, the suppression of vector charmonium $J/\psi$ was first proposed in Ref.~\cite{Matsui:1986js}.
Assuming that the QGP is created, the liberated quarks and gluons screen the color charges of heavy quarks like the Debye screening~\cite{Satz:1986ds,DeGrand:1986ds}.
As a consequence, the binding of heavy quark and antiquark pairs in the QGP can be weakened and thus the yield of quarkonia is suppressed.
With the temperature increasing further, the dissociation of quarkonia begins and the heavy quarks eventually diffuse in the QGP.
However, because of the nuclear shadowing effect, the Cronin effect, the nuclear absorption, and the multi-channel correlation (or level crossing),
the suppression may not come uniquely from the formation of QGP (see, e.g., Refs.~\cite{Qiu:1986,Cronin:1975,Huefner:1988a,Gu:1999,Karsch:2006}).
Some investigations show that, to understand the charmonium production at RHIC,
one should consider not only the suppression effect but also the regeneration process (see, e.g., Refs.~\cite{Zhuang:2006,Zhuang:2014,Mueller:2018}.
Moreover, comparing the $J/\psi$ yields observed in RHIC and LHC experiments (see, e.g., Refs.~\cite{Adamczyk:2012,Adam:2016}) with the summary of the yields of particle productions in LHC~\cite{Munzinger:2018}, one can recognize that there exists an abnormal enhancement of the $J/\psi$ production with respect to the particles (light nuclei) with similar mass, which deviates from the results of the statistical model dramatically.
The mechanism of the abnormal enhancement of the $J/\psi$ production needs to be clarified imperatively.

Due to the temperature dependence of the color screening radius, it is expected that mesons with different flavors may dissociate at different temperatures \cite{Digal:2001iu,Kaczmarek:2003dp,Mocsy:2008eg}.
This means that the in-medium mesons can serve as a probe of the QGP.
Recent results of lattice QCD show that the heavy quarkonia can survive above the critical temperature $T_{c}$ of the chiral symmetry restoration
and deconfinement~\cite{Aarts:2011cm,Umeda:2007cm,Umeda:2008cm,Karsch:2004cm,Asakawa:2003cm,Aarts:2007cm,Ohno:2011cm,Quinn:2019cm}.
However, light flavor mesons may dissociate in the neighborhood of $T_{c}$~\cite{Ding:2014oc,KLWang:2013,Hohler:2014rho}.
In other words, the properties of in-medium mesons are closely related to the chiral symmetry restoration and deconfinement.

The properties of in-medium mesons are encoded in the spectral functions of mesonic correlators \cite{LeBellac:2000wh}.
The difficulty is twofold, i.e., non-perturbative calculation of the correlators and reliable extraction of the spectral functions.
The lattice QCD is a first-principle non-perturbative approach to solve QCD.
Combining with the maximum entropy method (MEM) of the spectral analysis, it has obtained numerous interesting results,
such as, the evolution of bound state peaks with temperatures, heavy quark diffusion coefficients in the QGP, electrical conductivities of the QGP,
and so on~\cite{Aarts:2007cod,Ding:2012D1,Aarts:2013cod,Francis:2012cod,Ding:2012D2,Ding:2018D}.
Besides, many other approaches have also been applied in the studies (see, e.g., Refs.~\cite{KLWang:2013,Hohler:2014rho,Eichten:1980,Fu:2009,Fernandez:2009,Fujita:2009,Mocsy:2008eg,Jung:2016}).

The Dyson-Schwinger equations (DSEs), known as a continuum QCD approach including both dynamical chiral symmetry breaking (DCSB) and confinement,
have been successfully applied in studying hadron properties and QCD phase transitions (see, e.g., Refs.~\cite{all1,all2,all3,all4,all5,all6,all7,Zong:2013}).
In the DSE framework, in-vacuum hadrons are described by the bound state equations, e.g., the two-body Bethe-Salpeter equation (BSE) and the three-body Faddeev equation. By solving the equations, one can study in-vacuum hadrons properties (see, e.g., Refs.~\cite{Chang:2011ei,Blank:2011ha,Chang:2013nia,Chang:2013pq,Maris:2003vk,Chang:2011vu}).
However, at nonzero temperatures, the definition of bound state equations is problematic since bound states may dissociate. Moreover, the numerical procedures become complicated because the Matsubara frequencies are introduced in imaginary-time thermal field theory~\cite{LeBellac:2000wh}. Recently, a novel spectral representation has been successfully developed to extend the DSE approach for in-medium hadron properties and QGP transport properties~\cite{Qin:2015spf1, Qin:2014spf2}.

In this work, we study the flavor dependence of the dissociation temperatures of vector and axial-vector mesons in the DSE framework, by analyzing the behaviors of the corresponding spectral functions. We observe that the dissociation temperature increases distinctly with the ascending of the current quark mass. We shed light then on the abnormal increase of the $J/\psi$ yield obtained in LHC experiments.

The paper is organized as follows. In Section~\ref{RfMC}, we describe briefly the theoretical framework of the mesonic correlators.
In Section~\ref{SFs}, we depict the spectral representation and reconstruction.
In Section~\ref{RaD}, we present our numerical results and discussions.
Section~\ref{SaO} provides a summary and perspective.

\section{Mesonic correlators}
\label{RfMC}

According to the imaginary-time formalism of thermal field theory \cite{LeBellac:2000wh}, the mesonic correlators of a local operator $J_{H}(\tau,\vec{x}\,)$ is defined as
\begin{eqnarray}
G_H(\tau,\vec{x}\,)=\langle J_{H}(\tau,\vec{x}\,)
J_{H}^\dagger(0,\vec{0}\,) \rangle_{\beta} \, ,
\end{eqnarray}
where $\beta=1/T$, $\tau$ is the imaginary time with $0<\tau<\beta$,
and $\langle \ldots \rangle_\beta$ denotes the thermal average.
The operator $J_{H}$ has the following form
\begin{eqnarray}
J_{H}(\tau,\vec{x}\,)=\bar{q}(\tau,\vec{x}\,)\gamma_H q(\tau,\vec{x}\,)\;,
\end{eqnarray}
with $\gamma_{H} = \mathbf{1}$, $\gamma_{5}$, $\gamma_{\mu}$, $\gamma_{5}\gamma_{\mu}$
for scalar, pseudo-scalar, vector, axial-vector channel, respectively.

In terms of Green's functions, the Euclidean mesonic correlators are defined as
\begin{eqnarray}
\parbox{78mm}{\includegraphics[width=\linewidth]{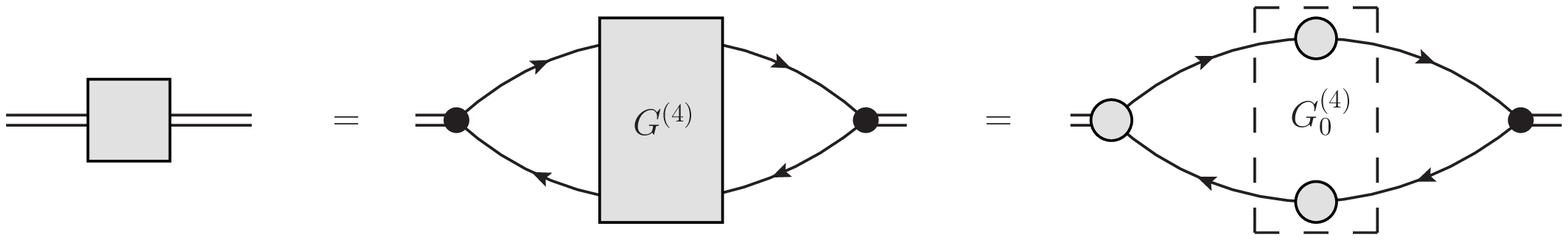}} \; , \; \; \;\;
\label{eq:cor}
\end{eqnarray}
where gray circular blobs denote dressed propagators $S$ and
vertices $\Gamma_H$, $G^{(4)}$ denotes the full quark--antiquark four-point Green's function,
$G_{0}^{(4)}$ stands for  the two disconnected dressed quark propagators in the dashed box, and
black dots refer to the bare propagators or vertices.
As the basic building blocks of the correlators, the dressed propagators $S$  and vertices $\Gamma_{H}$ in Eq.\ \eqref{eq:cor},
have to be solved self-consistently by the corresponding DSEs.

On one hand, the gap equation for the dressed quark propagator $S$ reads
\begin{eqnarray}
\parbox{76mm}{\includegraphics[width=\linewidth]{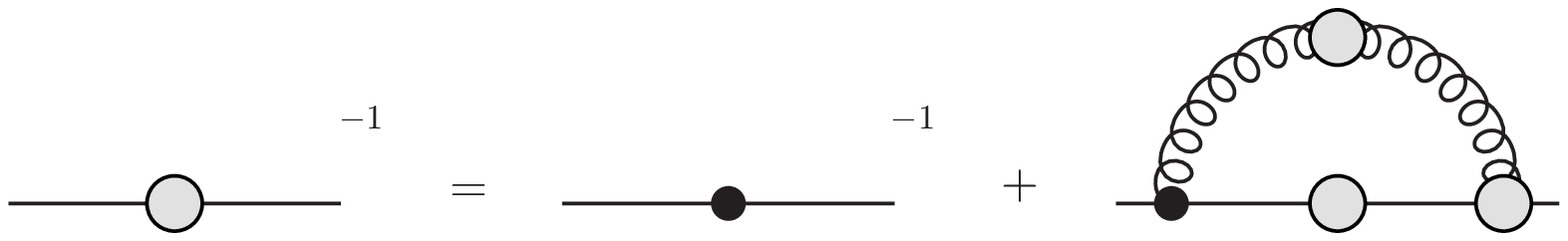}} \;. \;\;\;\;\;\;
\label{eq:gap}
\end{eqnarray}
From the above equation, it is found that $S$ depends on the dressed gluon propagator $D_{\mu\nu}^{ab}$ and the dressed quark-gluon vertex $\Gamma_{\mu}^{a}$, explicitly.
On the other hand, the dressed vertex $\Gamma_H$ satisfies the inhomogeneous BSE,
\begin{eqnarray}
\parbox{76mm}{\includegraphics[width=\linewidth]{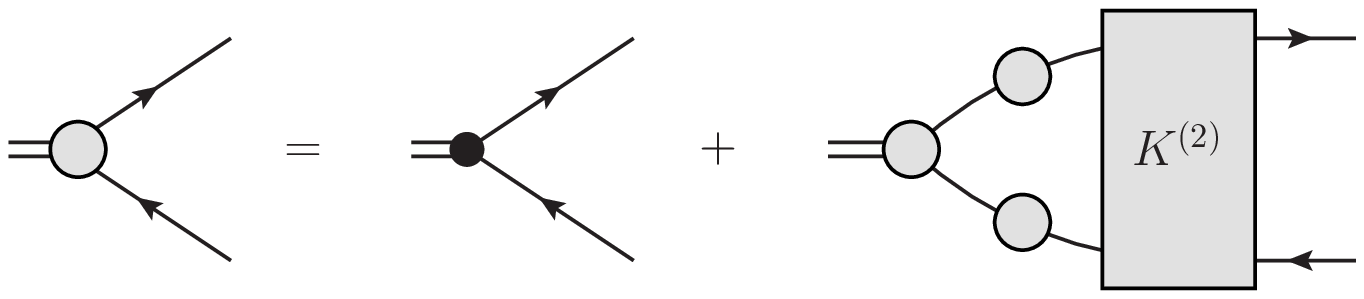}} \; , \;\;\;\;\;\;
\label{eq:bse}
\end{eqnarray}
where $K^{(2)}$ denotes the two-particle irreducible kernel and the dressed quark propagators are fed with the solutions of the gap equation. The solutions can be decomposed according to the $J^{P}$ quantum number of the corresponding channel $H$. To sum up, in order to solve Eqs.\ \eqref{eq:gap} and \eqref{eq:bse}, we have to specify the three objects $D_{\mu\nu}^{ab}$, $\Gamma_\mu^a$, and $K^{(2)}$.

To this end, we adopt the widely used rainbow-ladder (RL) approximation (see Ref.\cite{Eichmann:2008} and references therein), which is the leading symmetry-preserving scheme to satisfy the Ward-Takahashi identities \cite{Ball:1980ay, Maris:1997hd, Qin:2013mta, Qin:2014vya}. The rainbow part of this approximation is expressed as (color indices are suppressed)
\begin{equation}
Z_{1} g^{2} D_{\mu\nu}(k_{\Omega})\Gamma_{\nu}(\omega_{n},\vec{p}\,;\omega_{l},\vec{q}\,)
=D^{\rm eff}_{\mu\nu}(k_{\Omega})\gamma_{\nu}\;,
\end{equation}
with the effective gluon propagator written as
\begin{equation}
D^{\rm eff}_{\mu\nu}(k_{\Omega}) = P^{T}_{\mu\nu}\mathcal{D}(k_{\Omega}^{2})
+P^{L}_{\mu\nu}\mathcal{D}(k_{\Omega}^{2} + m_{g}^{2}) \, , \;
\end{equation}
where $k_{\Omega} = (\omega_{n} - \omega_{l},\, \vec{p}-\vec{q}\,)$,
$P^{T,L}_{\mu\nu}$ are the transverse and longitudinal projection tensors, respectively.
$\mathcal{D}$ is the gluon dress function which describes the effective interaction,
and the gluon Debye mass $m_{g}^{2} = (16/5)T^{2}$.
Consequently, the gap equation can be written explicitly as
\begin{eqnarray}
S(\omega_{n},\vec{p}\,)^{-1} &=& Z_2(i\vec{\gamma}\cdot\vec{p} + i\gamma_4 \omega_{n} + Z_m m)\notag\\
	&&+\, \frac{4T}{3} \sum_l\int\frac{d^3\vec{q}}{(2\pi)^3} D_{\mu\nu}^{\rm eff} (k_\Omega){\gamma_{\mu}} S(\omega_l,\vec{q}\,) \gamma_\nu\,,\, \quad
\label{eq:gap0}
\end{eqnarray}
where $\omega_l=(2l+1)\pi T, l\in Z$, are the fermionic Matsubara frequencies; $Z_{2,m}$ is the quark wave function and mass renormalization constants, respectively.
The solution $S(\omega_{n},\vec{p}\,)$ can be generally decomposed as
\begin{eqnarray}
S(\omega_n,\vec{p}\,)^{-1} &=& i\vec{\gamma}\cdot\vec{p}\,
A(\omega_n^2,\vec{p}\,^2) + i\gamma_4\omega_n\,
C(\omega_n^2,\vec{p}\,^2) \notag\\
&& +~ B(\omega_n^2,\vec{p}\,^2),
\end{eqnarray}
where $A,B$, and $C$ are scalar functions. The quark mass scale can be defined as $M(\vec{0},\omega_0^2) := B(\vec{0},\omega_0^2)/A(\vec{0},\omega_0^2)$,
which can be taken as an order parameter of the chiral phase transition.

The ladder part of the RL approximation expresses the two-particle irreducible kernel in terms of the one-gluon exchange form
\begin{eqnarray}
K^{(2)}(\omega_n,\vec{p}\,;\omega_l,\vec{q}\,)= -\frac{4}{3} D^{\rm eff}_{\mu\nu}(k_\Omega)(\gamma_\mu \otimes \gamma_\nu)\;,
\end{eqnarray}
Inserting the above expression into Eq. \eqref{eq:bse}, the inhomogeneous BSE can be rewritten as
\begin{eqnarray}
 && \Gamma_H(\omega_n;\omega_m,\vec{p}\,)=Z_H\gamma_H - \frac{4}{3}\sum_l\int\frac{d^3\vec{q}}{(2\pi)^3} {g^{2}} D_{\mu\nu}^{\rm eff} (k_\Omega) \notag\\
  &&\qquad\times\,{\gamma_{\mu}}S(\omega_l,\vec{q}\,)\Gamma_H(\omega_n;\omega_l,\vec{q}\,) S(\omega_l+\omega_n,\vec{q})\gamma_\nu,\quad
\label{eq:bse0}
\end{eqnarray}
where $\omega_{n} = 2n\pi T,\, n\in Z$, are the bosonic Matsubara frequencies;
and the renormalization constant $Z_{H}$ is, respectively, $Z_{4}$ ($=Z_{2} Z_{m}$) and $Z_{2}$ for the (pseudo-)scalar and the (axial-)vector.

Now the quark gap equation and the inhomogeneous BSE, i.e.,
Eqs.\ \eqref{eq:gap0} and \eqref{eq:bse0}, can be solved once the gluon dress function $\mathcal{D}$ is specified. Here, we adopt the one-loop renormalization-group-improved interaction model \cite{Qin:2011dd,Qin:2011xq}
\begin{equation}\label{QC}
\mathcal{D}(s)=\frac{8\pi^2}{\xi^4} \eta e^{-s/\xi^2}+\frac{8\pi^2\gamma_m\mathcal{F}(s)}{\ln[\tau+(1+s/\Lambda_{QCD}^2)^2]}\,,
\end{equation}
which has two parameters: the width $\xi$ and the strength $\eta$ with the product $\xi \eta$ characterizing the effective interaction strength. Generally, one can fix the parameters by fitting the properties of in-vacuum pseudoscalar mesons.
In this work, following Ref.~\cite{Qin:2018pp, Qin:2019hgk}, we take $\xi=0.5\,$GeV and $\xi=0.8\,$GeV for light and heavy sectors, respectively.

\section{Spectral functions}
\label{SFs}

The useful information is encoded in the mesonic spectral functions, which are related to the imaginary parts of the retarded correlators,
\begin{eqnarray}
\rho_H(\omega,\vec{p}\,)&=&2\,{\rm Im}\,G_H^R(\omega,\vec{p}\,)
=2\,{\rm Im}\,G_H(i\omega_n,\vec{p}\,)|_{ i\omega_n\to\omega+i\epsilon}\;.\notag\\
\label{eq:img}
\end{eqnarray}
Then, the spectral representation at zero momentum ($\vec{p}\,=\vec{0}\,$) reads
\begin{eqnarray}
G_H(\omega_n^{2})=\int_{0}^{\infty} \frac{d\omega^2}{2\pi}
\frac{\rho_H(\omega)}{\omega^2 +\omega_n^2}-({\rm
subtraction})\;,
\label{eq:spec1}
\end{eqnarray}
where an appropriate subtraction is necessary due to the divergence of spectral integral, i.e., $\rho_{H}(\omega\to\infty)\propto \omega^2$.
Following Refs.~\cite{Qin:2015spf1, Qin:2014spf2}, we introduce a discrete transformation for $G_{H}$ and define the transformed correlator as,
\begin{eqnarray}
\tilde G_{H}(\omega_{n}^{2}) &=&
\frac{G_{H}(\omega_{n}^{2})}{(\omega_{n}^{2}-\omega_{n+1}^{2})(\omega_{n}^{2}-\omega_{n+2}^{2})}\notag\\
&&+\frac{G_{H} (\omega_{n+1}^{2})}{(\omega_{n+1}^{2}-\omega_{n}^{2})(\omega_{n+1}^{2}-\omega_{n+2}^{2})}\notag\\
&&+\frac{G_{H} (\omega_{n+2}^{2})}{(\omega_{n+2}^{2}-\omega_{n}^{2})(\omega_{n+2}^{2}-\omega_{n+1}^{2})}\,.
\label{eq:transform}
\end{eqnarray}
Then, we have
\begin{equation}
\tilde G_{H}(\omega_{n}^{2})=\int_{0}^\infty\frac{d\omega^2}{2\pi}\frac{\rho_{H}(\omega)}{(\omega^2+\omega_n^2)(\omega^2+\omega_{n+1}^2)(\omega^2+\omega_{n+2}^2)}\,,\quad
\label{eq:spec3}
\end{equation}
which is divergence-free in both the ultraviolet and infrared regions.
Compared with the original expression Eq.~\eqref{eq:spec1}, the spectral representation in Eq.~\eqref{eq:spec3} can serve as the practical tool
for the extraction of observables.

The spectral representation connects the Euclidean correlator which can be calculated in the DSE framework with the spectral function which encodes observables.
However, it is generally an ill-posed problem to reconstruct the spectral function since its degrees of freedom are much more than the data points of the calculated correlator. Thus, a prior knowledge for the spectral functions is required.
To solve the problem, one can introduce an ansatz parameterizing the spectral function and fit the parameters by the standard $\chi^2$-procedure.

At zero temperature, since the ground state of a meson dominates the corresponding spectral integral, we can then parameterize the spectral function as
\begin{equation}
\rho_{H}(\omega)=C_{\rm res}\frac{M\Gamma\omega^2}{(\omega^2-M^2)^2+M^2\Gamma^2}+C_{\rm cut} \, \omega^2 \Theta(\omega^2-\omega_{0}^2)\,,
\label{eq:zerospf}
\end{equation}
where the first term is the Breit-Wigner distribution~\cite{Boon:1980,Forster:1990,Ding:2011ud} and
the second term is the simplified perturbative continuum branch cut~\cite{Altherr:1989tail}.
Inserting the above ansatz into the spectral representation, one can express the correlator in terms of the parameters.
The fitted parameters can give the interested observables, e.g., in-vacuum mass spectra.
In our calculations, we do not include explicitly the radial excitation states because their signals are very weak~\cite{Qin:2015spf1}.

 At nonzero temperature, we extend the ansatz as (see Ref. \cite{Ding:2018sto} for an example)
 \begin{eqnarray}
\rho_H(\omega)&=&C_{\rm trs}\frac{\eta \ \omega}{\eta^2+\omega^2}
+C_{\rm res}\frac{M \Gamma\omega^2}{(\omega^2-M^2)^2+M^2\Gamma^2}\notag\\
&& + ~ C_{\rm cut} \, \omega^2 \Theta(\omega^2-\omega_{0}^2),
\label{eq:finitespf}
\end{eqnarray}
where the first term is introduced for the transport peak. The properties of in-medium mesons can be read off from the evolution of the parameters with temperature, e.g., the mass and the width, which can signal the dissocaiation of mesons as we will see in the next section.

\section{Results and discussions}
\label{RaD}

\begin{table*}
\caption{Masses of some light falvor and heavy flavor mesons in vacuum. The results are obtained  by both the spectral function analysis (herein)
and solving the homogeneous BSE (h.BSE). Dimensional quantities are displayed in GeV.
\label{tab:1}}
\begin{tabular*}
{\hsize}
{|l@{\extracolsep{0ptplus1fil}}
|l@{\extracolsep{0ptplus1fil}}
l@{\extracolsep{0ptplus1fil}}
l@{\extracolsep{0ptplus1fil}}
l@{\extracolsep{0ptplus1fil}}
l@{\extracolsep{0ptplus1fil}}
l@{\extracolsep{0ptplus1fil}}
l@{\extracolsep{0ptplus1fil}}
l@{\extracolsep{0ptplus1fil}}
l@{\extracolsep{0ptplus1fil}}
l@{\extracolsep{0ptplus1fil}}
l|@{\extracolsep{0ptplus1fil}}}\hline
       & $\pi$  & $K$ & $\sigma$ & $\rho$ & $a_1$ & $\phi$  & $f_1$ & $\eta_c$ & $J/\psi$ & $\chi_{c0}$ & $\chi_{c1}$ ~    \\\hline
herein\; & $0.138$  & $0.493$ & $0.655$ & $0.771$ & $0.939$  & $1.09$ & $1.29$ & $3.01$ & $3.11$ & $3.45$ & $3.55$ \\
h.BSE & $0.138$ & $0.495$ & $0.660$ & $0.768$ & $0.921$ & $1.09$ & $1.24$ & $2.97$ & $3.09$ & $3.31$ & $3.44$   \\\hline
\end{tabular*}
\end{table*}

\begin{figure}
\centering
\includegraphics[width=0.46\textwidth]{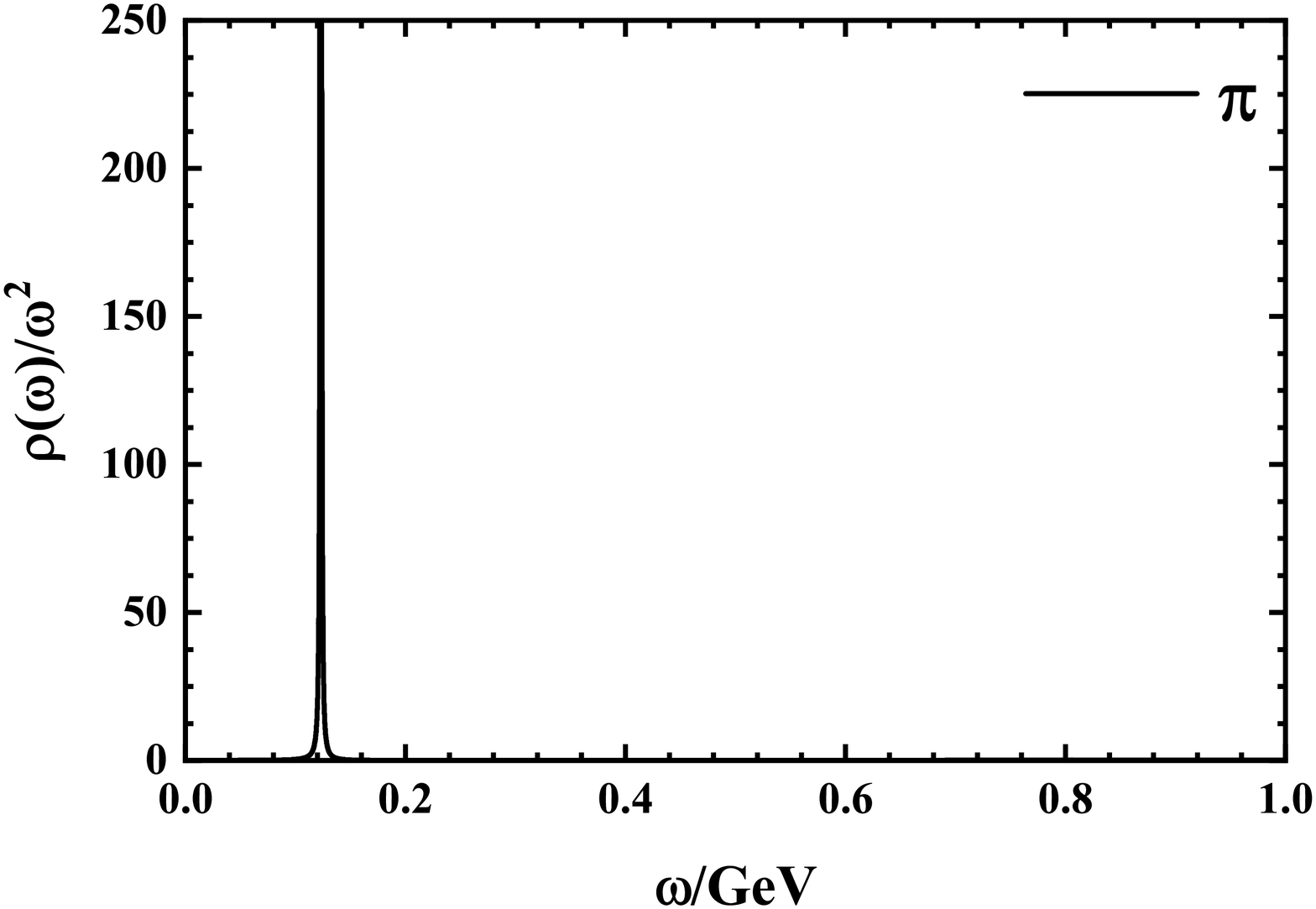} \hspace{6mm}
\includegraphics[width=0.445\textwidth]{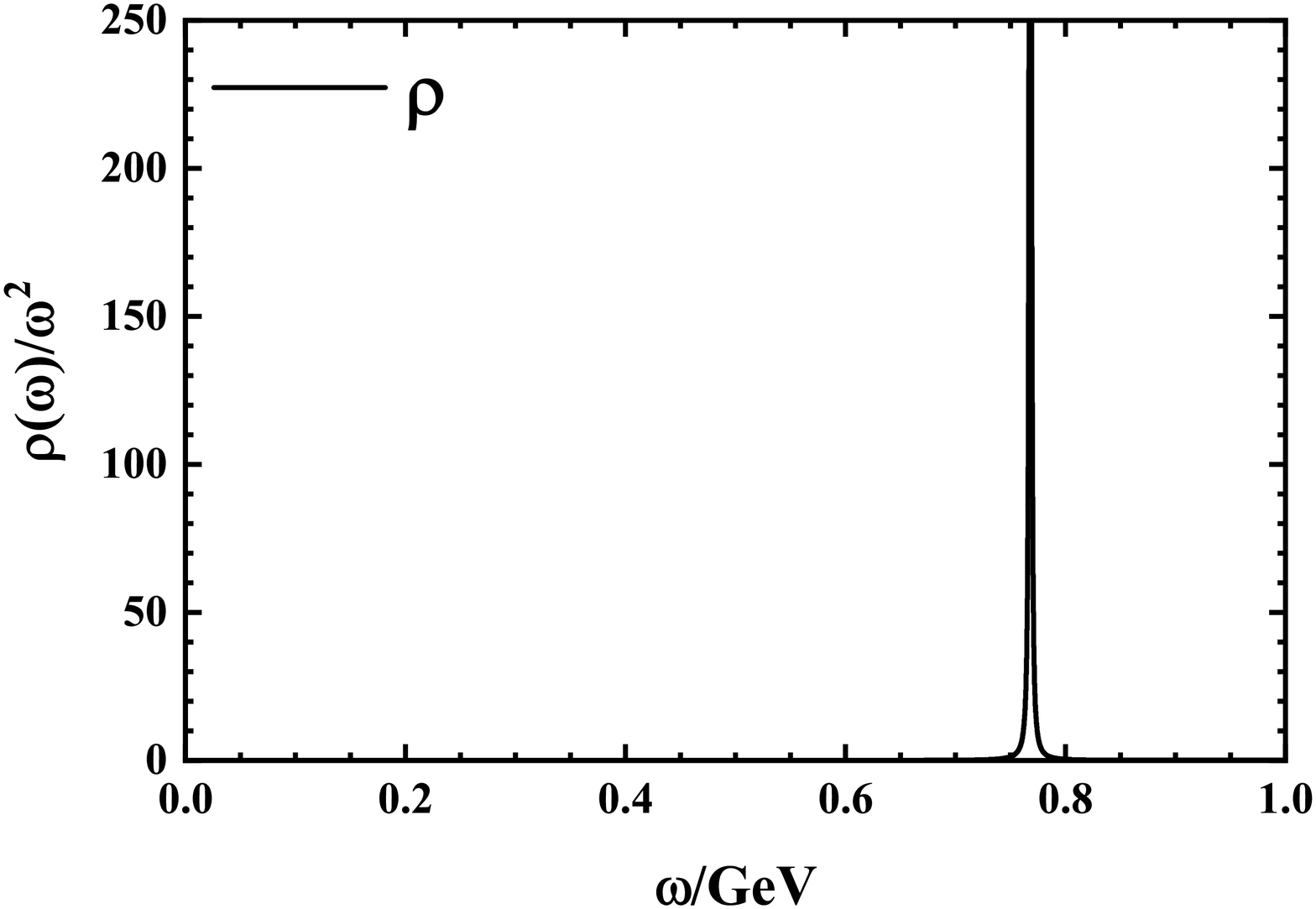}
\caption{Calculted pseudoscalar and vector mesons' spectral functions at zero temperature.}
\label{fig:zero}
\end{figure}

At zero temperature, the spectral functions have five parameters which can be fitted with a quite high precision.
For example, the fitted spectral functions of the pseudoscalar and vector mesons' spectral functions at zero temperature can be shown in Fig.~\ref{fig:zero}.
It can be found apparently that the Breit-Wigner peaks are very sharp, which means that the ground states correspond to the simple poles in the Green's functions.
For such a case, the homogeneous BSE works very well.
As a comparison, we also include the results obtained by solving the homogeneous BSE in Table~\ref{tab:1}.
It can be noted easily that the relative difference between the two methods is less than $5\%$.

Notice that, since the ground state peaks are very sharp and strong, the signals of excited states are so weak which is beyond our fitting accuracy.
Thus, one could turn to more sophisticated numerical techniques, e.g., the maximum entropy method (MEM)~\cite{Asakawa:2000mem}.
However, it is still very difficult to obtain robust results for excited states~\cite{Qin:2015spf1}.
Nevertheless, the studies on excited states are beyond the scope of this work.
Next, we will focus on the properties of the ground state mesons at nonzero temperature.

\begin{figure}
\centering
\includegraphics[width=0.46\textwidth]{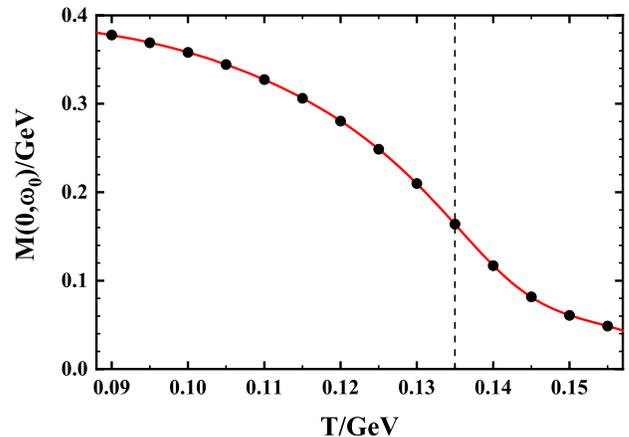}\hspace{6mm}
\caption{Calculated evolution feature of the light flavor quark mass scale with temperatures, where the dashed vertical line indicates the steepest descent temperature.}
\label{fig:Tc}
\end{figure}

At nonzero temperature, we first study the evolution of the light quark mass scale with temperatures.
The obtained result is displayed in Fig.~\ref{fig:Tc}.
It is evident that, with temperature increasing, the mass scale gradually decreases.
Moreover, there is a temperature region where the decrease becomes very rapid.
The steepest descent temperature is usually taken to define the (pseudo)critical temperature $T_{c,\chi}$ of the chiral phase transition,
which is indicated by the dashed vertical line in the figure.
We have then $T_{c,\chi}^{(l)}=135\,$MeV, which is consistent with the state of the art lattice QCD simulation~\cite{Ding:2019prx}.

\begin{figure}
\centering
\includegraphics[width=0.44\textwidth]{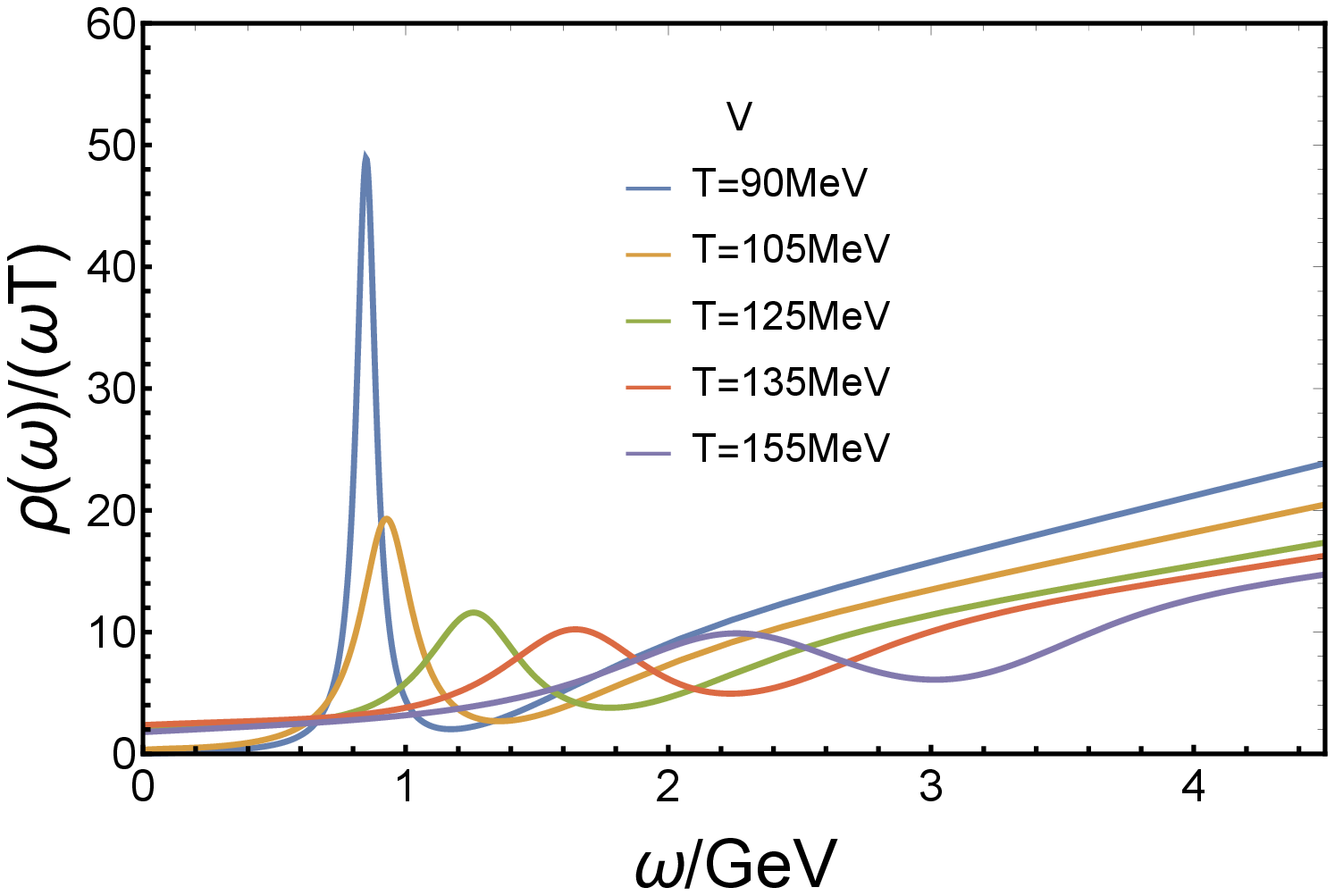}\hspace{10mm}
\includegraphics[width=0.44\textwidth]{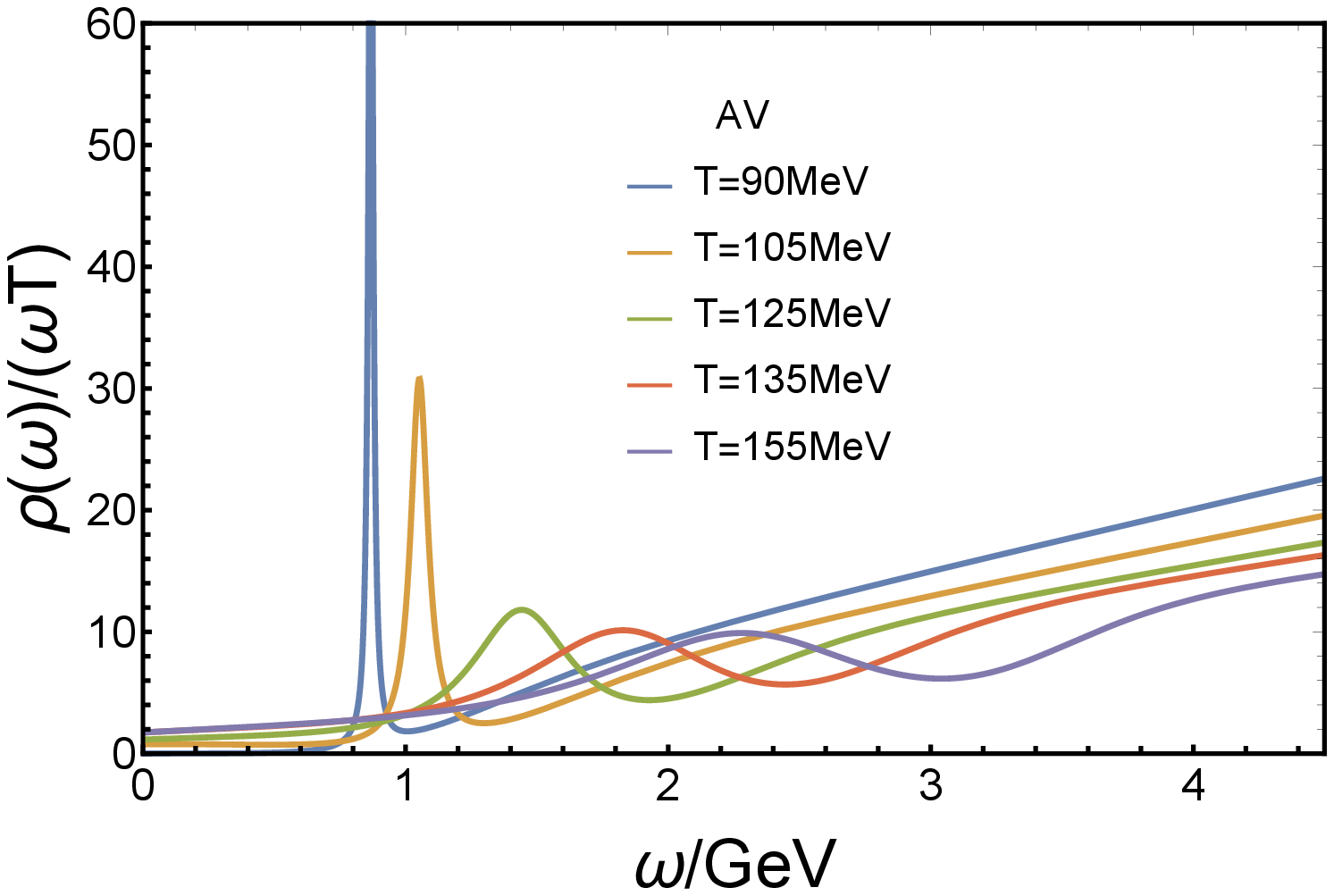}\hspace{10mm}
\caption{Spectral functions of the light quark vector (upper panel) and axial-vector (lower panel) channels obtained by fitting the corresponding correlators at several temperatures: $90\sim 155$ MeV.} \label{fig:spec}
\end{figure}

Next we analyze the spectral functions of the light quark vector and axial-vector channels at different temperatures.
Our calculated spectral functions are shown in Fig.~\ref{fig:spec}, where the results for temperatures below 90~MeV are skipped since there is no significant change
from that at zero temperature. It is easily found that, with temperature increasing, the ground state peaks become broad and decrease in height.
Especially, in the neighborhood of the $T_{c,\chi}^{(l)}$, the peaks are dramatically smeared and eventually become indistinguishable from the background.
This means that, at high temperature, the bound states may dissociate.

\begin{figure}
\centering
\includegraphics[width=0.44\textwidth]{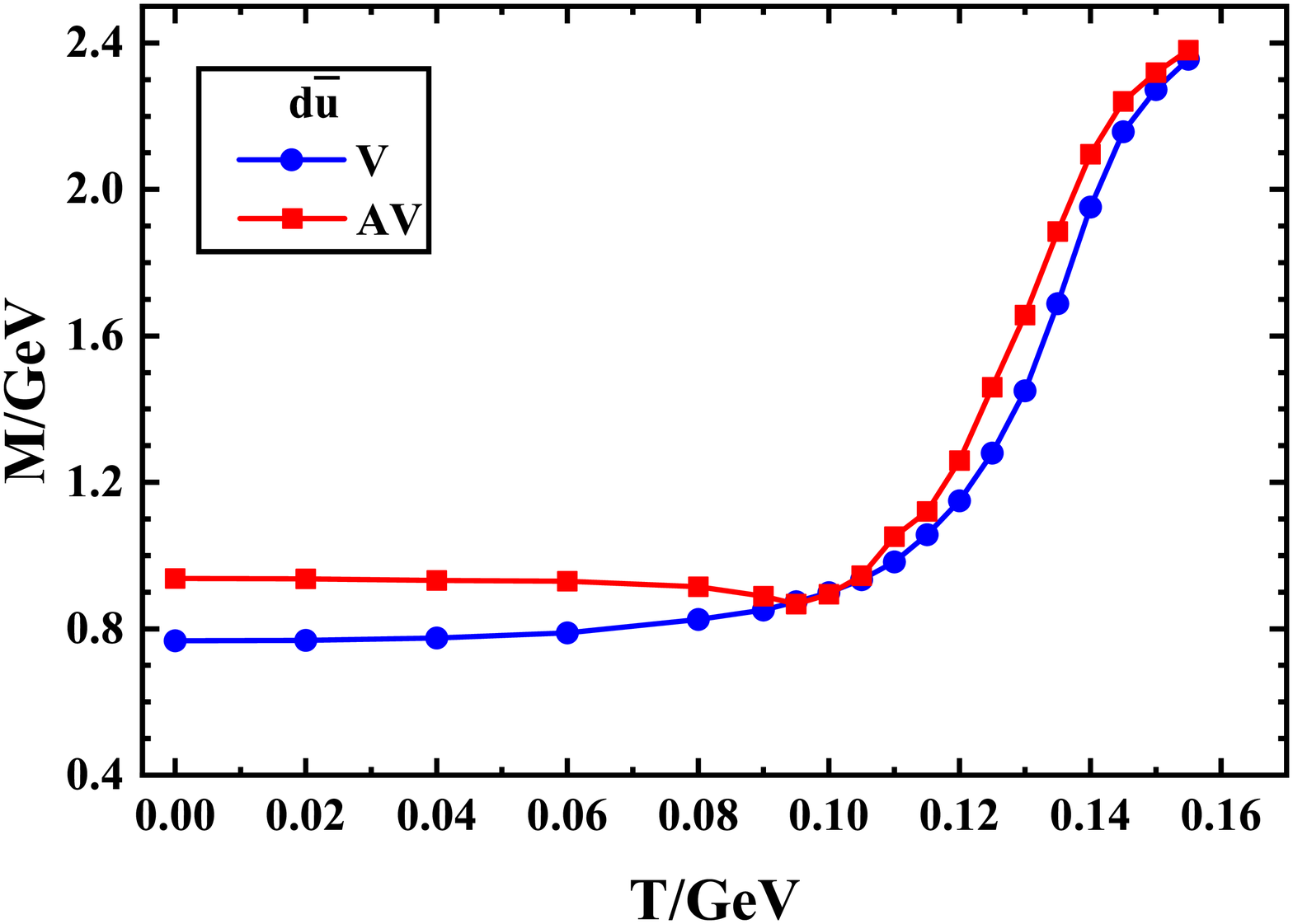}\hspace{6mm}
\includegraphics[width=0.44\textwidth]{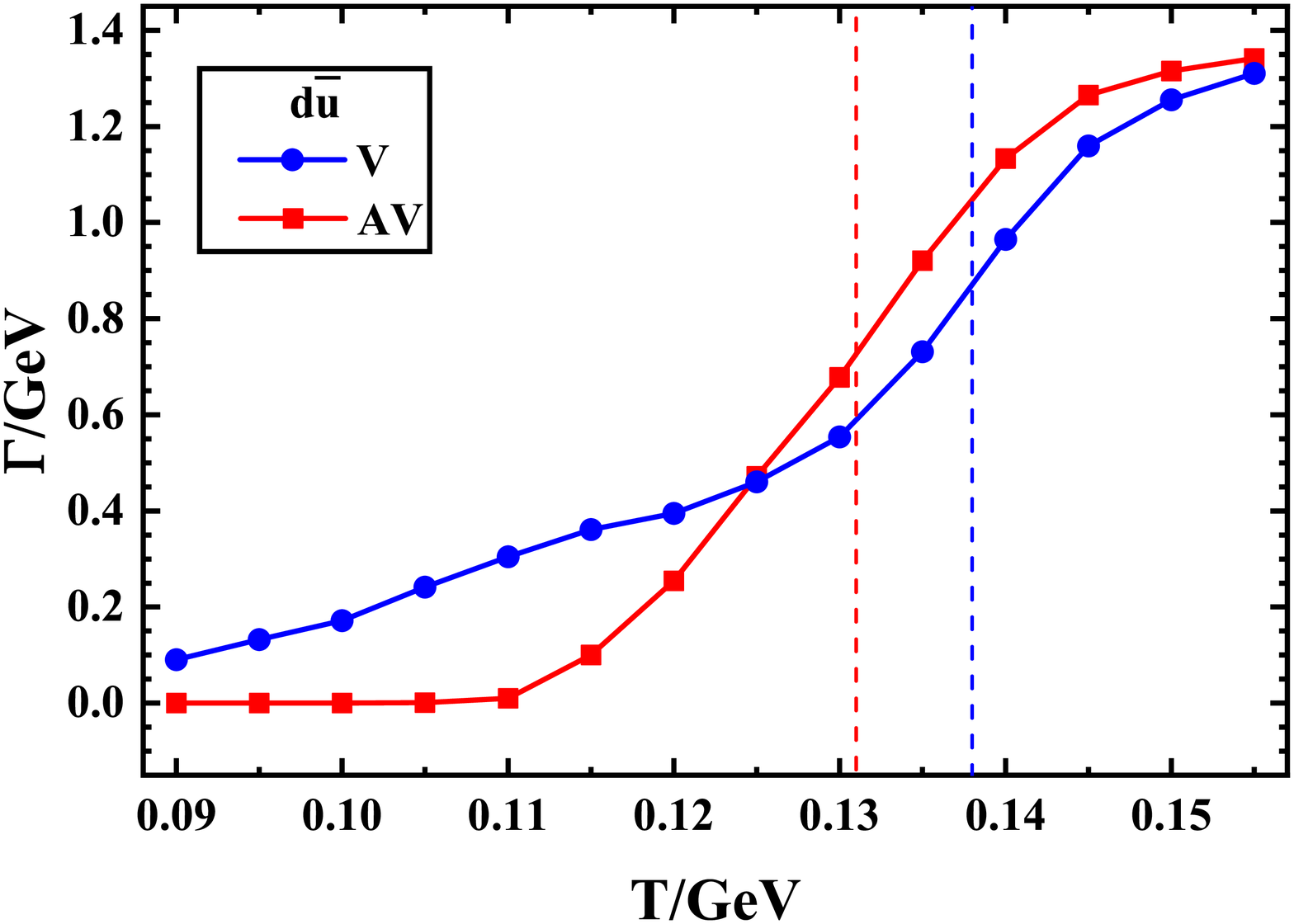}\hspace{6mm}
\caption{Calculated temperature dependence of the masses and widths of the peaks in the light quark vector and axial-vector channels, where the dashed vertical lines indicates the steepest ascent temperatures.}
\label{fig:massgamma}
\end{figure}

In order to further understand the dissociation process, we focus on the features of the peaks, i.e., the masses $M$ and the widths $\Gamma$.
The obtained results are shown in Fig.~\ref{fig:massgamma}.
One can observe evidently from Fig.~\ref{fig:massgamma} that, the masses remain almost unchanged until $T\sim 110\,$MeV and dramatically increase
when $T\sim T_{c,\chi}^{(l)}$.
At very high temperatures, the difference between the masses of the vector and axial-vector channels become invisible.
It can also be easily noticed that the masses of the two channels accidentally coincide at $T\sim 100$ MeV.
This roots in the drawback of the RL approximation which significantly underestimates the mass of the light quark axial-vector channel due to the lack of
enough spin-orbital repulsion~\cite{Chang:2011ei,Qin:2011dd,Qin:2011xq}.
If taking the reason into consideration, the axial-vector mass should be much larger than the presented result in the temperature region below $T_{c,\chi}^{(l)}$, and the accidental coincidence could disappear.

Fig.~\ref{fig:massgamma} illustrates also obviously that, the widths remain small at low temperatures and increase rapidly at $T\sim T_{c,\chi}^{(l)}$.
Similar to analyzing the quark mass scale, we can study the steepest ascent temperatures of the widths, denoted by $T_{s}$.
We have that $T_{s}\sim 138\,$MeV for the vector channel and $T_s\sim 131\,$MeV for the axial-vector one (see the vertical lines in Fig.~\ref{fig:massgamma}).
Above the steepest ascent temperatures, the widths become comparable with the masses so that the peaks can hardly be identified as bound states.
Since $T_{s} \sim T_{c,\chi}^{(l)}$ with only several MeV difference, it is sound to conclude that the dissociation of light quark bound states happens
at the (pseudo)critical temperature of the chiral phase transition.

\begin{figure*}
\centering
\includegraphics[width=0.44\textwidth]{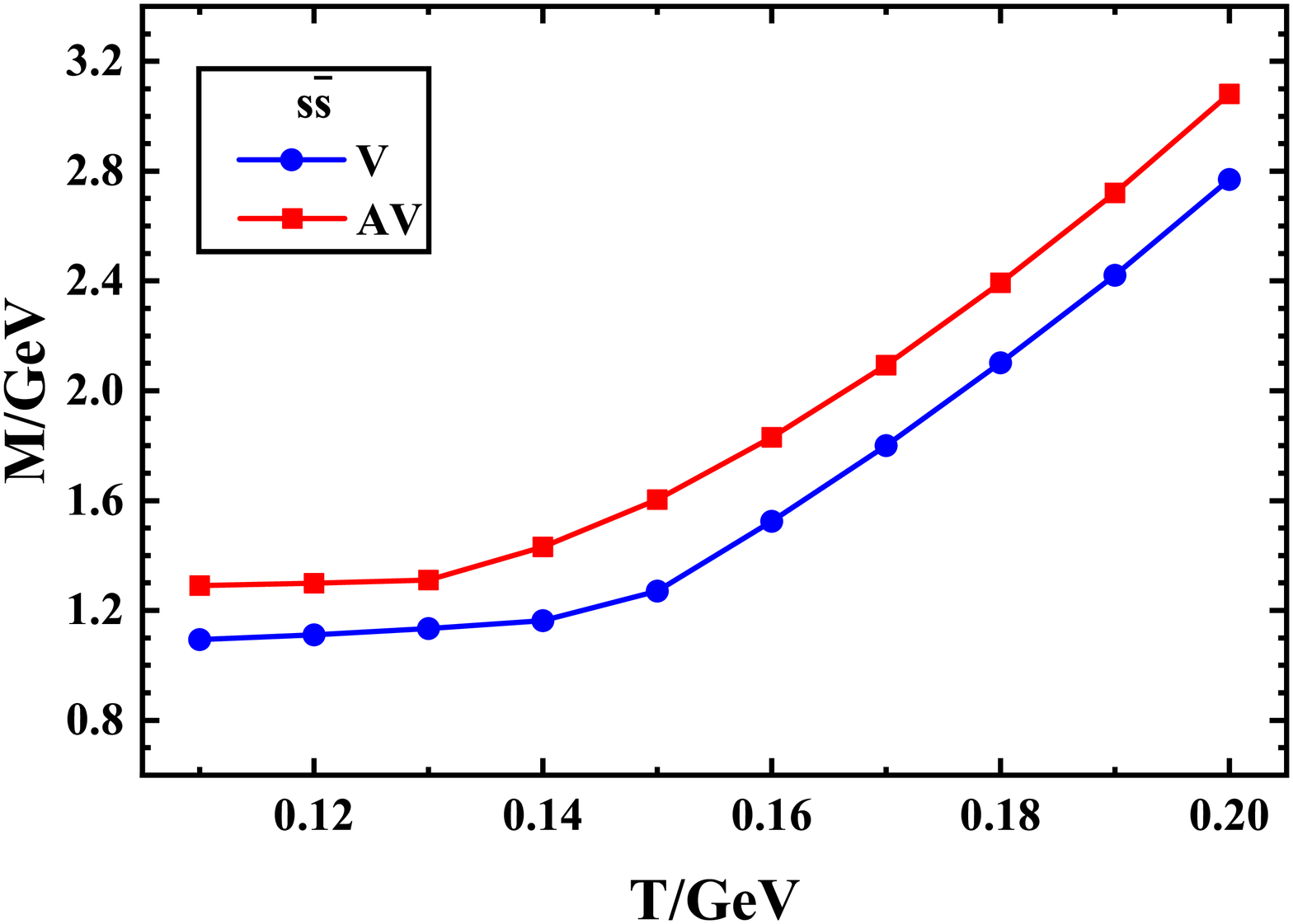}  \hspace*{5mm}
\includegraphics[width=0.44\textwidth]{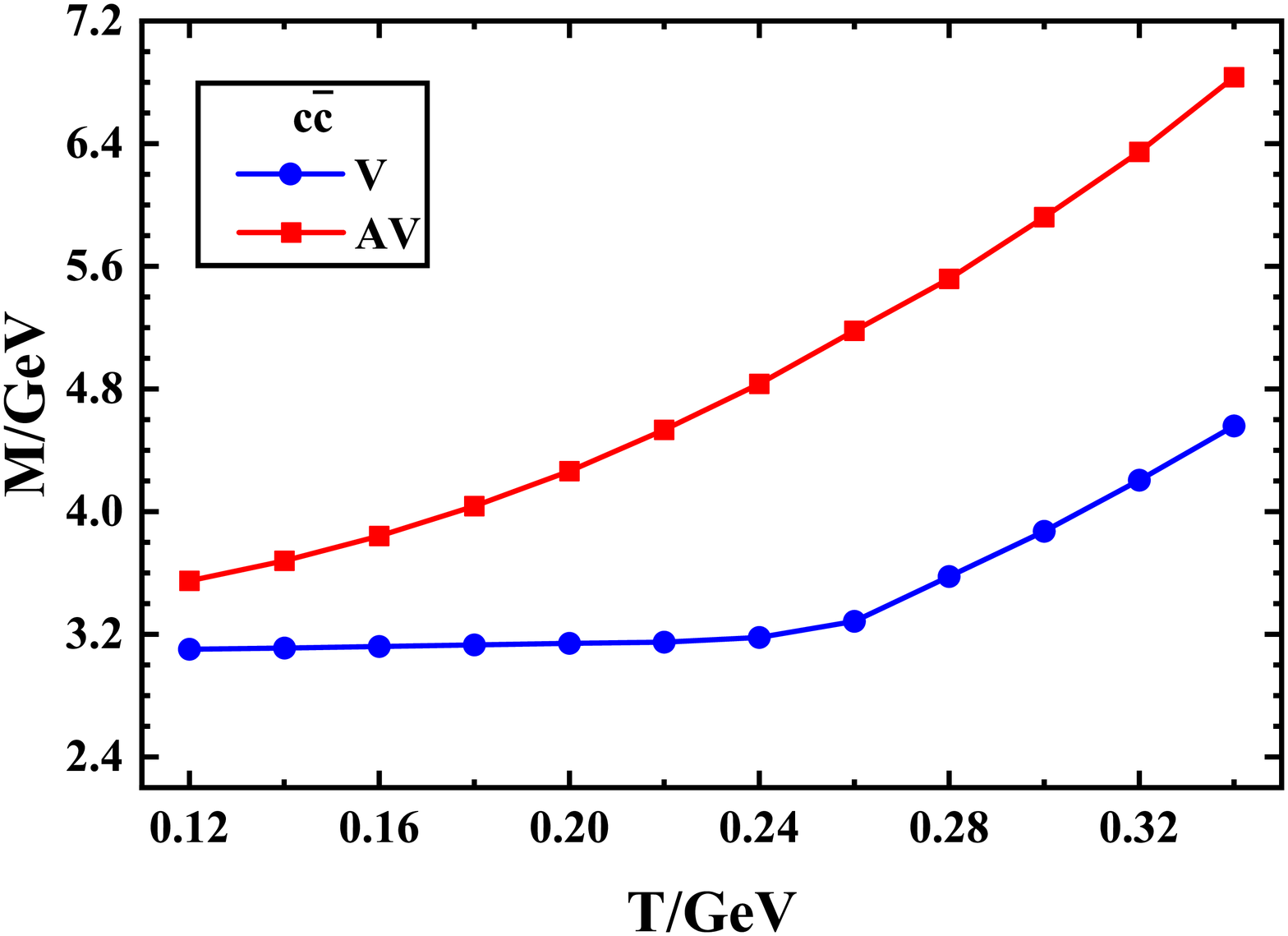}
\includegraphics[width=0.44\textwidth]{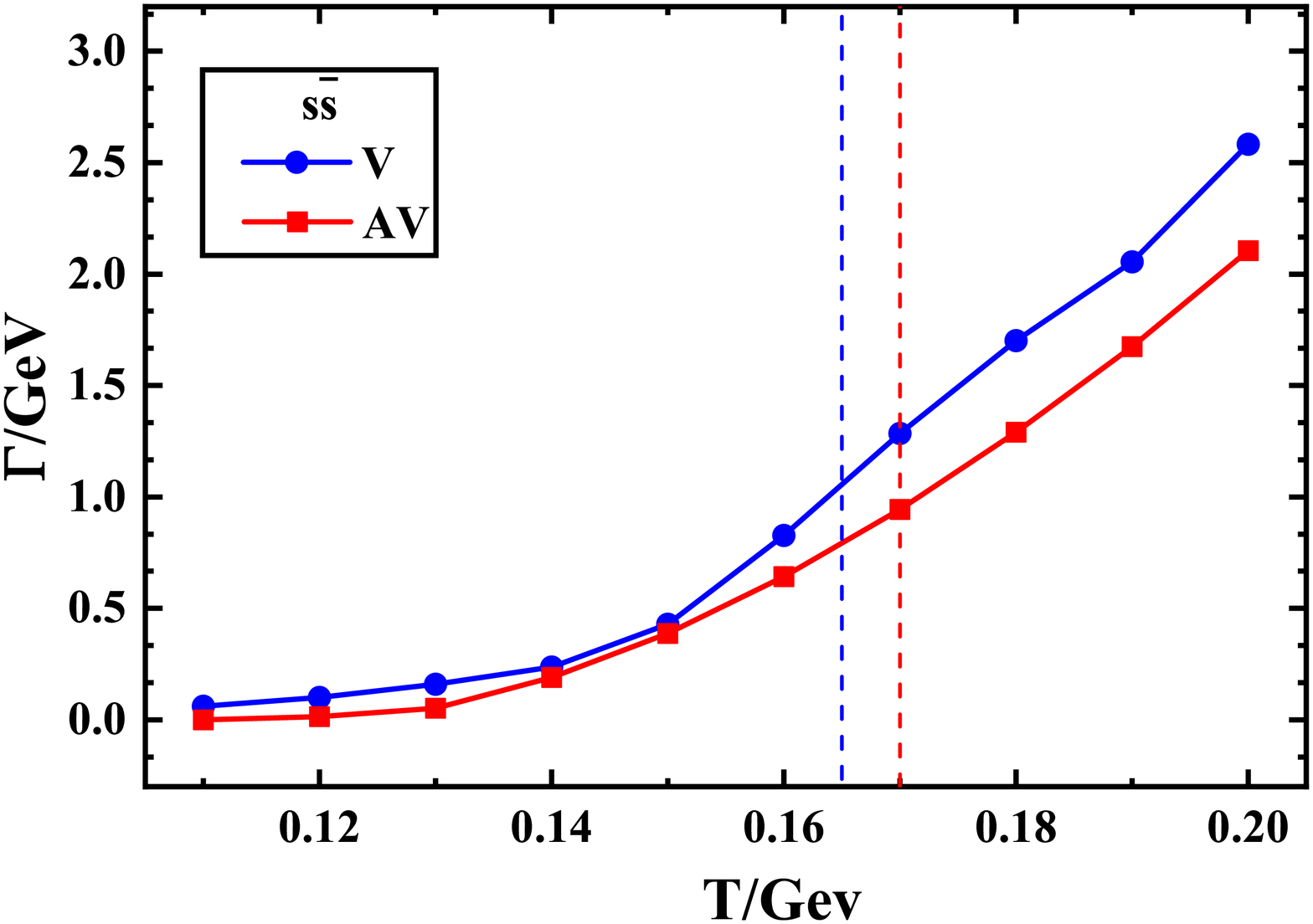} \hspace*{5mm}
\includegraphics[width=0.44\textwidth]{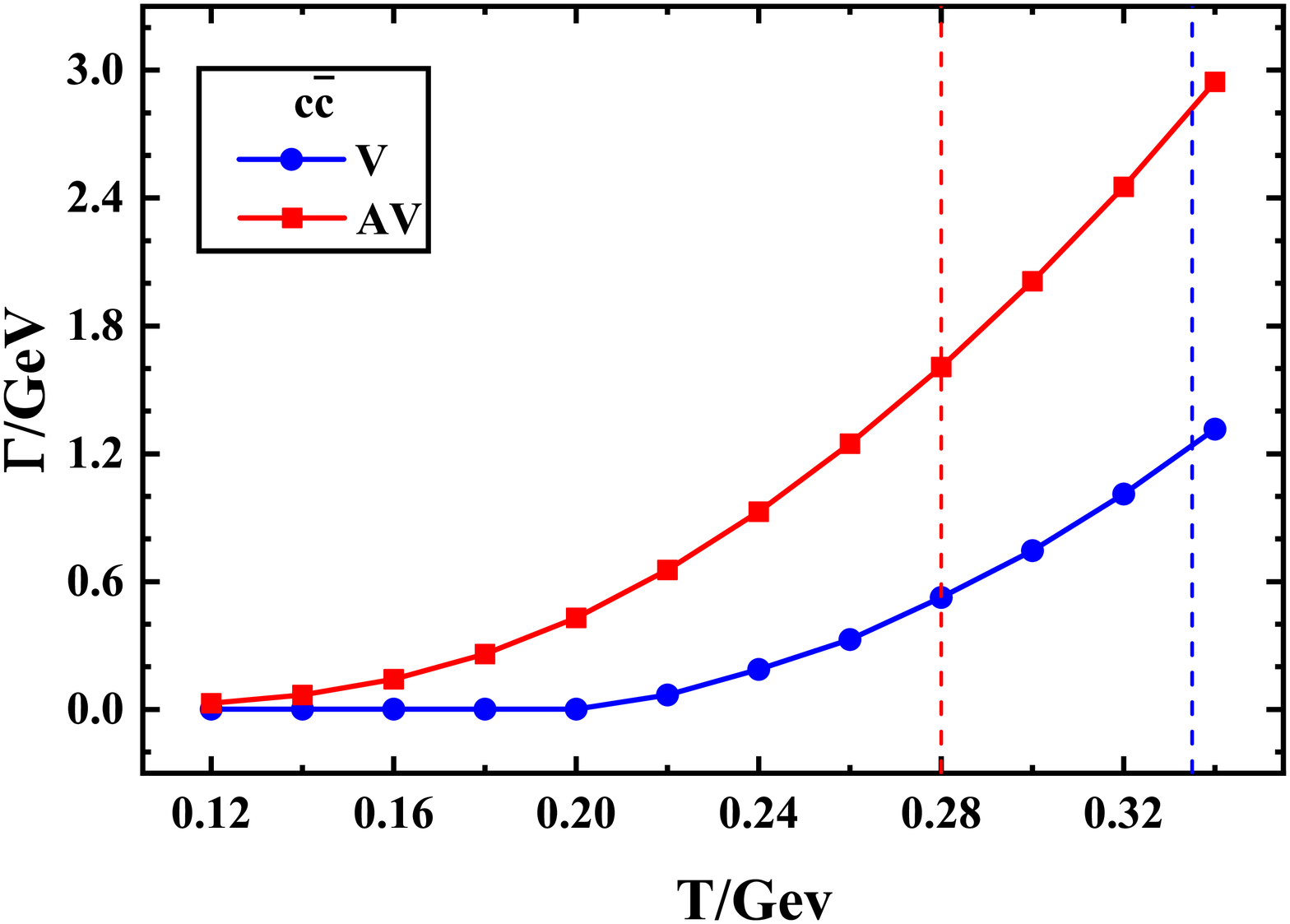}
\caption{Calculated temperature dependence of the masses and widths of the peaks in the strange (left panel) and charm (right panel) quark vector and axial-vector channels, where the dashed vertical lines manifests the steepest ascent temperatures.} \label{fig:sc}
\end{figure*}

Now a natural question arises: can any bound states survive at high temperatures. In order to address the question, we study the dependence of the spectral functions on the quark flavors. We first analyze the evolution of masses of strange and charm quark bound states, e.g., $s\bar{s}$ and $c\bar{c}$, with temperatures. The results are shown in Fig. \ref{fig:sc}. Unlike the light flavor case, it is found that the pole masses of vector and axial-vector channels do not coincide at high temperatures any more. It is much more obvious in $c\bar{c}$ bound states which should be understood as that the dominant contribution comes from the explicit chiral symmetry breaking rather than DCSB in the heavier flavor cases~\cite{all2}.

From Fig. \ref{fig:sc}, one can also recognize that, unlike the cases of light flavor quarks, the masses and the widths of $s\bar{s}$ and $c\bar{c}$
remain almost unchanged in the neighborhood of $T_{c,\chi}^{(l)}$.
For $s\bar{s}$ vector and axial-vector mesons, the properties start being changed by the thermal environment for $T>150\;$MeV,
and the steepest ascent temperatures of the widths, i.e., the dissociation temperatures $T_{s}\sim170\;$MeV.
For $c\bar{c}$ cases, the thermal effect becomes visible at very high temperatures, i.e., $T > 2 T_{c,\chi}^{(l)}$.
The dissociation temperatures $T_{s} \sim 280\;$MeV and $330\;$MeV for vector and axial-vector $c\bar{c}$ mesons, respectively.
This observation can be regarded as a flavor dependence of the dissociation and may intuitively explain the flavor hierarchy in the deconfinement transition~\cite{Bellwied:2013fla}.
Moreover, similar results can be found in other works using different methods (see, e.g., Refs. \cite{Bazavov:2015,Shi:2013,Brandt:2016char,Cheng:2011,Karsch:2004cm,Quinn:2019cm}).
Such a result also means that the formation temperature of heavy flavor mesons from quark gluon matter is much higher than that for light flavor mesons.
In turn, the formation phase space of heavy flavor mesons is very large.
Therefore the yield of the $J/\psi$ in the LHC experiments should be much larger than light flavor particles with the similar mass (e.g., the light nuclei $^{3}\textrm{H}$,
${^{3}_{\Lambda}}\textrm{He}$, etc.).

\section{Summary and outlook}
\label{SaO}
In summary, based on the spectral representation of mesonic correlators and the self-consistent solutions of the rainbow-ladder truncated BSE and DSE,
we studied systematically the temperature dependence of the meson spectral functions in both vector and axial-vector channels,
especially, the masses and the widths of the spectral peaks.
The results reveal the flavor dependent pattern of the thermal dissociations of vector and axial-vector mesons.
For light flavor mesons, the masses and the widths dramatically increase in the crossover region of the chiral restoration.
It is sound to conclude that the dissociation of light flavor mesons and the chiral symmetry restoration happen, simultaneously.
However, for strange and charm quark mesons, the dissociation temperatures increase, significantly.
Moreover, the heavier the quarks, the higher the dissociation temperatures.

As having mentioned in the context, the RL approximation can work well only for several channels. Then sophisticated truncation scheme beyond the RL approximation,
such as with the full quark-gluon vertex~\cite{Qin:2013mta,Chang:2011},  should be taken in the spectral function analysis,
and an interaction model with more realistic temperature dependence could also be adopted.
Moreover, the studies at not only nonzero temperature but also nonzero chemical potentials are of great interest,
more specifically, the region in the vicinity of the possible critical end point.
The related works are under progress.

\begin{acknowledgments}
The work was supported by the National Natural Science Foundation of China under Contracts No. 11435001, No. 11775041, No. 11805024, No.11947406.
%
%
\end{acknowledgments}


\begin{thebibliography}{99}

\bibitem{phenix:2008jexp}
 A. Adare {\it et al.}, PHENIX Collaboration,
\newblock Phys. Rev. Lett.  {\bf 101}, 122301 (2008).

\bibitem{sci:2012}
B.V. Jacak and B. Muller,
\newblock Science  {\bf 337}, 310 (2012).

\bibitem{Aarts:2017}
G. Aarts {\it et al.},
\newblock Eur. Phys. J. A {\bf 53}, 93 (2017).

\bibitem{Matsui:1986js}
T. Matsui and H. Satz,
\newblock Phys. Lett. B {\bf 178}, 416 (1986).

\bibitem{Satz:1986ds}
K. Kanaya and H. Satz,
\newblock Phys. Rev. D {\bf 34}, 3193 (1986).

\bibitem{DeGrand:1986ds}
T.A. DeGrand and C.E. DeTar,
\newblock Phys. Rev. D {\bf 34}, 2469 (1986).

\bibitem{Qiu:1986}
A. H. Mueller, and J. W. Qiu,
\newblock Nucl. Phys. B  {\bf 268}, 427 (1986).

\bibitem{Cronin:1975}
J. W. Cronin, H. J. Frisch, M. J. Shochet, J. P. Boymond, P. A. Pirou\'{e}, and R. L. Sumner,
Phys. Rev. D {\bf 11}, 3105 (1975);
J. Huefner, Y. Kurihara, and H. J. Pirner,
\newblock Phys. Lett. B {\bf 215}, 218 (1988).

\bibitem{Huefner:1988a}
C. Gerschel, and J. Huefner,
\newblock Phys. Lett. B {\bf 207}, 253 (1988).

\bibitem{Gu:1999}
J. Z. Gu, H. S. Zong, Y. X. Liu, and E. G. Zhao,
\newblock Phys. Rev. C {\bf 60}, 035211 (1999).

\bibitem{Karsch:2006}
F. Karsch, D. Kharzeev, and H. Satz,
\newblock Phys. Lett. B {\bf 637}, 75 (2006).

\bibitem{Zhuang:2006}
L. Yan, P. Zhuang, and N. Xu,
\newblock Phys. Rev. Lett. {\bf 97}, 232301 (2006).

\bibitem{Zhuang:2014}
K. Zhou, N. Xu, Z. Xu, and P. Zhuang,
\newblock Phys. Rev. C {\bf 89}, 054911 (2014).

\bibitem{Mueller:2018}
X. Yao, and B. Mueller,
\newblock Phys. Rev. C {\bf 97}, 014908 (2018).

\bibitem{Adamczyk:2012}
L. Adamczyk, et al., STAR Collaboration,
\newblock Phys. Rev. C {\bf 86}, 024906 (2012).

\bibitem{Adam:2016}
J. Adam, et al., ALICE Collaboration,
Phys. Rev. Lett. {\bf 116}, 222301 (2016).

\bibitem{Munzinger:2018}
A. Andronic, P. Braun-Munzinger, K. Redlich, and J. Stachel,
Nature {\bf 561}, 321 (2018).

\bibitem{Digal:2001iu}
S.~Digal, P.~Petreczky and H.~Satz,
\newblock Phys.\ Lett.\ B {\bf 514}, 57 (2001);

\bibitem{Kaczmarek:2003dp}
O. Kaczmarek, F. Karsch, P. Petreczky, and F. Zantow
\newblock Nucl. Phys. Proc. Suppl. 129, 560 (2004);

\bibitem{Mocsy:2008eg}
A.~Mocsy,
\newblock Eur.\ Phys.\ J.\ C {\bf 61}, 705 (2009).

\bibitem{Asakawa:2003cm}
M. Asakawa and T. Hatsuda,
\newblock Phys. Rev. Lett.  {\bf 92}, 012001 (2004).

\bibitem{Karsch:2004cm}
S. Datta, F. Karsch, P. Petreczky, and I. Wetzorke,
\newblock Phys. Rev. D {\bf 69}, 094507 (2004).

\bibitem{Umeda:2007cm}
T. Umeda,
\newblock Phys. Rev. D {\bf 75}, 094502 (2007).

\bibitem{Aarts:2007cm}
G. Aarts, C. Allton, M.B. Oktay, M. Peardon, and J.-I. Skullerud,
\newblock Phys. Rev. D {\bf 76}, 094513 (2007).

\bibitem{Umeda:2008cm}
H. Ohno, T. Umeda, and K. Kanaya,
\newblock  J. Phys. G {\bf 36}, 064027 (2009).

\bibitem{Ohno:2011cm}
H. Ohno {\it et al.}, WHOT-QCD Collaboration,
\newblock Phys. Rev. D {\bf 84}, 094504 (2011).

\bibitem{Aarts:2011cm}
G. Aarts, C. Allton, S. Kim, M.P. Lombardo, M.B. Oktay, S.M. Ryan, D.K. Sinclair, and J.I. Skullerud,
\newblock JHEP, {\bf 11}, 103 (2011).

\bibitem{Quinn:2019cm}
R. Quinn, J. Glesaaen, A. Rothkopf, and J. Skullerud,
\newblock PoS Confinement2018, 272 (2019).

\bibitem{Ding:2014oc}
A. Bazavov {\it et al.},
\newblock Phys. Lett. B {\bf 737}, 210 (2014).

\bibitem{KLWang:2013}
K.L. Wang, Y.X. Liu, L. Chang, C.D. Roberts, and S.M. Schmidt,
\newblock Phys. Rev. D {\bf 87}, 074038 (2013).

\bibitem{Hohler:2014rho}
P.M. Hohler and R. Rapp,
\newblock Phys. Lett. B {\bf 731}, 103-109 (2014).

\bibitem{LeBellac:2000wh}
M. Le Bellac,
\newblock {\it {Thermal Field Theory}}, Cambridge University Press, 2000.

\bibitem{Aarts:2007cod}
G. Aarts, C. Allton, J. Foley, and S. Hands,
\newblock Phys. Rev. Lett.  {\bf 99}, 022002 (2007).

\bibitem{Ding:2012D1}
H.T. Ding, A. Francis, O. Kaczmarek, F. Karsch, H. Satz, and W. Soldner,
\newblock EPJ Web Conf. 70, 00061 (2014).

\bibitem{Aarts:2013cod}
A. Amato, G. Aarts, C. Allton, P. Giudice, S. Hands, and J. Skullerud,
\newblock Phys. Rev. Lett.  {\bf 111}, 172001 (2013).

\bibitem{Francis:2012cod}
A. Francis and O. Kaczmarek,
\newblock Prog. Part. Nucl. Phys.  {\bf 67}, 212-217 (2012).

\bibitem{Ding:2012D2}
H.T. Ding, A. Francis, O. Kaczmarek, F. Karsch, H. Satz, and W. Soeldner,
\newblock Phys. Rev.  D {\bf 86}, 014509 (2012).

\bibitem{Ding:2018D}
H.T. Ding, O. Kaczmarek, S. Mukherjee, H. Ohno, and H.T. Shu,
\newblock Phys. Rev. D {\bf 97}, 094503 (2018).

\bibitem{Eichten:1980}
E. Eichten, K. Gottfried, T. Kinoshita, K.D. Lane, and T.-M. Yan,
\newblock Phys. Rev. D {\bf 21}, 203 (1980).

\bibitem{Fu:2009}
W.J. Fu, and Y.X. Liu,
\newblock Phys. Rev. D {\bf 79}, 074011 (2009).

%
\bibitem{Fernandez:2009}
D.F. Fraile and A.G. Nicola,
\newblock Eur. Phys. J. C {\bf 62}, 37-54 (2009).

\bibitem{Fujita:2009}
M. Fujita, K. Fukushima, T. Kikuchi, T. Misumi, and M. Murata,
\newblock Phys. Rev. D {\bf 81}, 065024 (2010).

\bibitem{Jung:2016}
C. Jung, F. Rennecke, R.A. Tripolt, L. Smekal, and J. Wambach
\newblock Phys. Rev.  D {\bf 95}, 036020 (2016).

\bibitem{all1}
C.D. Roberts and A.G. Williams,
\newblock Prog. Part. Nucl. Phys. {\bf 33}, 477 (1994);
C.D. Roberts and S. Schmidt,
\newblock Prog. Part. Nucl. Phys. {\bf 45}, S1 (2000);
A. Bashir, L. Chang, I.C. Cloet, B. El-Bennich, Y.-X. Liu, C.D. Roberts, and P.C. Tandy,
\newblock Commun. Theor. Phys. {\bf 58}, 79 (2012).

\bibitem{all2}
L. Chang, Y. X. Liu, M. S. Bhagwat, C. D. Robert, and S. W. Wright,
\newblock Phys. Rev. C {\bf 75}, 015201 (2006);
K.L. Wang, S.X. Qin, Y.X. Liu, L. Chang, C.D. Roberts, and S.M. Schmidt,
\newblock Phys. Rev. D {\bf 86 }, 114001(2012); 
R. Williams, C. S. Fischer, and M. R. Pennington, 
\newblock Phys. Lett. B {\bf 645}, 167 (2007); 
C. S. Fischer, D. Nickel, and R. Williams, 
\newblock Phys. Lett. B {\bf 718}, 1036 (2013).

\bibitem{all3}
A. Bender, D. Blaschke, Y. Kalinovsky, and C.D. Roberts,
\newblock Phys. Rev. Lett. {\bf 77}, 3724 (1996);
P. Maris, C.D. Roberts, S.M. Schmidt, and P.C. Tandy,
\newblock Phys. Rev. C {\bf 63}, 025202 (2001).

\bibitem{all4}
W. Yuan, H. Chen, and Y. X. Liu, 
Phys. Lett. B {\bf 637}, 69 (2006);
S.X. Qin, L. Chang, H. Chen, Y.X. Liu, and C.D. Roberts,
\newblock Phys. Rev. Lett. {\bf 106}, 172301 (2011);
F. Gao, J. Chen, Y.X. Liu, S.X. Qin, C.D. Roberts, and S.M. Schmidt,
\newblock Phys. Rev. D {\bf 93}, 094019 (2016).

\bibitem{all5}
S.X. Qin, L. Chang, Y.X. Liu, and C.D. Roberts,
\newblock Phys. Rev. D {\bf 84}, 014017 (2011);
S.X. Qin, and D.H. Rischke,
\newblock Phys. Rev. D {\bf 88}, 056007 (2013).

\bibitem{all6}
C.S. Fischer, J. Luecker, and J.A. Mueller,
\newblock Phys. Lett. B {\bf 702}, 438 (2011);
C.S. Fischer and J. Luecker,
\newblock Phys. Lett. B {\bf 718}, 1036 (2013);
G. Eichmann, C.S. Fischer, and C.A. Welzbacher,
\newblock Phys. Rev. D {\bf 93}, 034013 (2016).

\bibitem{all7}
C.S. Fischer, L. Fister, J. Luecker, and J.M. Pawlowski,
\newblock Phys. Lett. {\bf B732}, 273 (2014);
C.S. Fischer, J. Luecker, and C.A. Welzbacher,
\newblock Phys. Rev. D {\bf 90}, 034022 (2014).

\bibitem{Zong:2013}
Z. F. Cui, C. Shi, Y. H. Xia, Y. Jiang, and H. S. Zong, 
\newblock Eur. Phys. J. C {\bf 73}, 2612 (2013);
S. S. Xu, Z. F. Cui, A. Sun, and H. S. Zong, 
J. Phys. G {\bf 45}, 105001 (2018).

\bibitem{Chang:2011ei}
L. Chang and C.D. Roberts,
\newblock Phys. Rev. C {\bf 85}, 052201 (2012).

\bibitem{Blank:2011ha}
M. Blank and A. Krassnigg,
\newblock Phys. Rev. D {\bf 84}, 096014 (2011).

\bibitem{Chang:2013nia}
L. Chang, I.C. Cloet, C.D. Roberts, S. Schmidt, and P. Tandy,
\newblock Phys. Rev. Lett. {\bf 111}, 141802 (2013).

\bibitem{Chang:2013pq}
L. Chang, I.C. Cloet, J.J. Cobos-Martinez, C.D. Roberts, S.M. Schmidt, and P.C. Tandy,
\newblock Phys. Rev. Lett.  {\bf 110}, 132001 (2013).

\bibitem{Maris:2003vk}
P. Maris and C.D. Roberts,
\newblock Int. J. Mod. Phys. E {\bf 12}, 297 (2003).

\bibitem{Chang:2011vu}
L. Chang, C.D. Roberts, and P.C. Tandy,
\newblock Chin. J. Phys.  {\bf 49}, 955 (2011).

\bibitem{Qin:2015spf1}
S.X. Qin,
\newblock Phys. Lett. B {\bf 742}, 358-362 (2015).

\bibitem{Qin:2014spf2}
S.X. Qin and D.H. Rischke,
\newblock Phys. Lett. B {\bf 734}, 157-161 (2014).

\bibitem{Eichmann:2008}
G. Eichmann, R. Alkofer, I. Cloet, A. Krassnigg, and C. Roberts,
\newblock Phys. Rev. C {\bf 77}, 042202 (2008).

\bibitem{Ball:1980ay}
J.S. Ball and T.-W. Chiu,
\newblock Phys. Rev. D {\bf 22}, 2542 (1980).

\bibitem{Maris:1997hd}
P.~Maris, C.~D.~Roberts and P.~C.~Tandy,
\newblock Phys.\ Lett.\ B {\bf 420}, 267 (1998).

\bibitem{Qin:2013mta}
S.X. Qin, L. Chang, Y.X. Liu, C.D. Roberts, and S.M. Schmidt,
\newblock Phys. Lett. B {\bf 722}, 384 (2013).

\bibitem{Qin:2014vya}
S.X. Qin, C.D. Roberts, and S.M. Schmidt,
\newblock  Phys. Lett. B {\bf 733}, 202 (2014).

\bibitem{Qin:2011dd}
S.X. Qin, L. Chang, Y.X. Liu, C.D. Roberts, and D.J. Wilson,
\newblock Phys. Rev. C {\bf 84}, 042202(R) (2011).

\bibitem{Qin:2011xq}
S.X. Qin, L. Chang, Y.X. Liu, C.D. Roberts, and D.J. Wilson,
\newblock Phys. Rev. C {\bf 85}, 035202 (2012).

\bibitem{Qin:2018pp}
S.X. Qin, C.D. Roberts, and S.M. Schmidt,
\newblock Phys. Rev. D {\bf 97}, 114017 (2018).

\bibitem{Qin:2019hgk}
S.~X.~Qin, C.~D.~Roberts and S.~M.~Schmidt,
\newblock Few Body Syst.\  {\bf 60}, no. 2, 26 (2019).

\bibitem{Boon:1980}
J.P. Boon and S. Yip,
\newblock  {\it Molecular Hydrodynamics}, McGraw-Hill, New York, (1980).

\bibitem{Forster:1990}
D. Forster,
\newblock {\it Hydrodynamics, Fluctuations, Broken Symmetry, and Correlation Functions}, Perseus-Books, Cambridge, MA, (1990).

\bibitem{Ding:2011ud}
H.T. Ding, A. Francis, O. Kaczmarek , F. Karsch, E. Laermann, and W. Soeldner,
\newblock Phys. Rev. D {\bf 83}, 033404 (2011).


\bibitem{Altherr:1989tail}
T. Altherr and P. Aurenche,
\newblock Z. Phys.  C {\bf 45}, 99 (1989).

\bibitem{Ding:2018sto}
H.T. Ding, O. Kaczmarek, S. Mukherjee, H. Ohno, and H.T Shu,
\newblock Phys. Rev. D {\bf 97}, 094503 (2018).

\bibitem{Asakawa:2000mem}
M. Asakawa, T. Hatsuda, and Y. Nakahara,
\newblock Prog. Part. Nucl. Phys.  {\bf 46}, 459 (2001).

\bibitem{Ding:2019prx}
H.T. Ding {\it et al.},
\newblock Phys. Rev. Lett. \textbf{123}, 062002 (2019).


\bibitem{Bellwied:2013fla}
R. Bellwied, S. Borsanyi, Z. Fodor, S. Katz, and C. Ratti
\newblock Phys. Rev. Lett.  {\bf 111}, 202302 (2013).

\bibitem{Brandt:2016char}
B.B. Brandt, A. Francis, B. Jager, and H. B. Meyer,
\newblock Phys. Rev. D {\bf 93}, 054510 (2016).

\bibitem{Cheng:2011}
M. Cheng, {\it et al.},
\newblock Eur. Phys. J. C {\bf 71}, 1564 (2011).

\bibitem{Shi:2013}
S. Shi, X. Guo, and P. Zhuang
\newblock Phys. Rev. D {\bf 88}, 014021 (2013).

\bibitem{Bazavov:2015}
A. Bazavov, F. Karsch, Y. Maezawa, S. Mukherjee, and P. Petreczky,
\newblock Phys. Rev. D {\bf 91}, 054503 (2015).

\bibitem{Chang:2011}
L. Chang, Y.X. Liu, and C.D. Roberts,
\newblock Phys. Rev. Lett. {\bf 106}, 072001 (2011);
%
C. Tang, F. Gao, and Y.X. Liu,
\newblock Phys. Rev. D {\bf 100}, 056001.


\end{thebibliography}
\end{document}